\documentclass[12pt,twocolumn]{openjournal}

\usepackage{xcolor}
\usepackage{textgreek}
\usepackage[utf8]{inputenc}
\usepackage[english]{babel}
\usepackage{hyperref}
\hypersetup{
    colorlinks=true,
    linkcolor=blue,
    filecolor=blue,      
    urlcolor=blue,
    citecolor=blue,
}
\usepackage{color,colortbl}
\usepackage{tensind}
\tensordelimiter{?}
\DeclareGraphicsExtensions{.bmp,.png,.jpg,.pdf}
\usepackage{needspace}
\usepackage{verbatim}
\usepackage[normalem]{ulem}
\usepackage{orcidlink}
\usepackage{soul}
\usepackage{xspace}
\usepackage{booktabs}

\begin{document}

\title{On the statistical convergence of N-body simulations of the Solar System}

\author{Hanno Rein\orcidlink{0000-0003-1927-731X}}

\affiliation{Dept. of Physical and Environmental Sciences, University of Toronto at Scarborough, Toronto, Ontario, M1C 1A4, Canada}
\affiliation{Dept. of Physics, University of Toronto, Toronto, Ontario, M5S 3H4, Canada}
\affiliation{Dept. of Astronomy and Astrophysics, University of Toronto, Toronto, Ontario, M5S 3H4, Canada}
\author{Garett Brown\orcidlink{0000-0002-9354-3551}}
\affiliation{Dept. of Physical and Environmental Sciences, University of Toronto at Scarborough, Toronto, Ontario, M1C 1A4, Canada}
\affiliation{Dept. of Physics, University of Toronto, Toronto, Ontario, M5S 3H4, Canada}

\author{Mei Kanda}
\affiliation{Dept. of Physical and Environmental Sciences, University of Toronto at Scarborough, Toronto, Ontario, M1C 1A4, Canada\\}

\shortauthors{Rein et al.}

\begin{abstract}
    Most direct N-body integrations of planetary systems use a symplectic integrator with a fixed timestep. 
    A large timestep is desirable in order to speed up the numerical simulations.
    However, simulations yield unphysical results if the timestep is too large.
    Surprisingly, no systematic convergence study has been performed on long (Gyr) timescales.
    In this paper we present numerical experiments to determine the minimum timestep one has to use in long-term integrations of the Solar System in order to recover the system's fundamental secular frequencies and instability rate.
    We find that timesteps of up to 32~days, i.e.\ a third of Mercury's orbital period, yield physical results in an ensemble of 5~Gyr integrations.
    We argue that the chaotic diffusion that drives the Solar System's long-term evolution dominates over numerical diffusion and timestep resonances.
    Our results bolster confidence that the statistical results of most simulations in the literature are indeed physical and provide guidance on how to run time and energy efficient simulations while making sure results can be trusted.
\end{abstract}

\maketitle
\section{Introduction}\label{sec:intro}
Analytical solutions and more recently numerical integrations of the Solar System, have a long history \citep[for a historical summary of the field see e.g.][]{Laskar2012}.
Different authors use different numerical algorithms, slightly different initial conditions, and different timesteps.
Reassuringly, there is general agreement about the main conclusion: the Solar System is marginally stable.
Specifically, there is a roughly~1\% chance that planets will collide with each other or get ejected within 5~Gyr, approximately the time that the sun will remain on the main sequence \citep{LaskarGastineau2009, Laskar2012, BrownRein2020,Abbot2021}.

Most modern simulations use symplectic integrators that make use of the so called Wisdom-Holman splitting \citep{Wisdom1981, WisdomHolman1991, Kinoshita1991}.
Various implementations of the Wisdom-Holman algorithm exist, with differences in the order, the coordinate system, and other details such as whether symplectic correctors are used or not.
For a review of splitting methods commonly used for planetary systems see \cite{ReinTamayoBrown2019}.
What all the methods have in common is that they require the user to choose a fixed timestep at the beginning of the simulation\footnote{The timestep is sometimes changed during the integration. For example it can be decreased when the eccentricity of a planet becomes large \citep{LaskarGastineau2009}. Although this formally breaks the symplecticity of the integrator it is in general not considered a problem to decrease the timestep. However, a later increase of the timestep would result in an increased error. Most importantly, the timestep should not be decreased and then increased repeatedly as the errors accumulate and the integrator looses all advantages symplectic integrators otherwise provide.}.
Timesteps from 3 days up to 9 days are typically used in simulations that include the inner Solar System.

Numerical convergence studies have been performed for short term integrations \citep{Rauch1999,Wisdom2015,Boekholt2015,Hernandez2022}.
In particular, \cite{Boekholt2015} show in several small $N$ test cases that ensembles of low accuracy integrations can be statistically indistinguishable from high accuracy ones when the energy error and the timestep are sufficiently small.
In a follow up, \cite{PortegiesZwart2018} study reversibility and strategies for obtaining converged solutions for unstable few-body problems.
The simulations of both \cite{Boekholt2015} and \cite{PortegiesZwart2018} were integrated for up to a few hundred dynamical timescales.
\cite{Trani2024} use a similar approach to study the three body problem over $10^{8}$ dynamical timescales.
They find that the phase space consists of regular and chaotic regions characterized by multi-fractal behaviour.
They furthermore show that numerical chaos can artificially enhance the mixing of these phase space regions and affect the reliability of individual simulations but preserve the statistical results of an ensemble of integrations.
But somewhat surprisingly a systematic convergence study has to our knowledge not been published for simulations of the full Solar System that are typically being integrated for billions of years, corresponding to $10^{10}$ dynamical timescales.

In this paper, we present such a statistical convergence study.
In contrast to other studies on symplectic integration methods, we do \emph{not} rely on conserved quantities such as the conservation of energy or angular momentum to assess convergence.
Whereas these quantities can give valuable hints regarding the performance of a numerical scheme, they are not interesting from an astrophysical point of view.
What we do instead is look at statistical quantities we actually care about from a physical point of view. 
In our case, these are a) the fundamental secular frequencies of the Solar System which are responsible for the evolution of the system over long timescales, and b) the ultimate stability of the system.

\Needspace{5\baselineskip}
\section{Secular Dynamics}
\label{sec:secdyn}
To better interpret our numerical results, let us provide a very brief summary of the dynamics of the Solar System that has emerged in recent years \citep[see e.g.][]{Hoang2021, Hoang2022, Mogavero2022}.

The Solar System is chaotic with a Lyapunov time of only 5~Myr \citep{Laskar1989}.
This implies that we can only predict the precise locations of planets over short timescales, $\lesssim 100$~Myr. 
To make predictions about the evolution of the Solar System over longer timescales, we have to integrate an ensemble of systems in which each has slightly different initial conditions.
These perturbations are typically of the order of centimetres and are thus all consistent with present day observations.
Initially the differences grow exponentially on the Lyapunov timescale.
Once the differences within the ensemble become macroscopic, the Solar systems enters a regime of chaotic diffusion. 
There are no (strong) mean-motion resonances present in the Solar System.
The diffusion is primarily driven by several secular resonances \citep{Laskar1990,LithwickWu2011}.
As orbital elements of planets diffuse over time, in particular those of the inner Solar System, they can enter an area of parameter space where close encounters or physical collisions are possible. 
The probability of such a catastrophic event is about 1\% \citep{LaskarGastineau2009}.
When an instability occurs, we typically stop the integration.

\Needspace{5\baselineskip}
\section{Numerical Setup}
\label{sec:setup}
We next describe our numerical setup.
It is similar to that in other studies where the goal is to determine the Solar System's instability rate with direct N-body integrations.

Our simulations consist of the sun and all eight planets with present day initial conditions taken from the NASA Horizons system.
The initial positions of Mercury are slightly perturbed in each run between $1$cm and $1$m.
We use \texttt{REBOUND} \citep{ReinLiu2012} and the Wisdom-Holman implementation \texttt{WHFast} \citep{ReinTamayo2015}, specifically the high order integrator \texttt{WHCKL} \citep{ReinTamayoBrown2019} with symplectic correctors and a modified kernel.
This method has a generalized order of $O(\epsilon\, dt^{18}+\epsilon^2 dt^{4} + \epsilon^3 dt^{3})$ where $\epsilon\sim 10^{-3}$ for the Solar System \citep{Wisdom1996}.
We include general relativistic corrections in the form of a modified solar potential \citep{Tamayo2020}.
We ignore asteroids, satellites, tides, higher order gravitational harmonics, radiation effects, and stellar mass loss.
For simplicity, we combine the Earth and Moon into one object at their barycenter.
We note that for realistic motions, the effects of the Moon should be taken into account \citep[e.g.][]{LaskarGastineau2009}.
We label a system as unstable when a planet gets ejected past 1000~AU or when two planets come closer than 4 times the Earth-Moon distance which is approximately the Hill radius of the Earth, about $1.5\cdot10^9$~m.
We choose this distance because when two planets come this close, their orbits will be significantly perturbed.
Although not guaranteed, there is a high probability of a physical collision or an ejection afterwards.
We note that our goal was not to resolve the individual collision or ejection but rather the point at which the system has been significantly altered.

Because our goal is to perform a statistical convergence study, we run 1200 simulations for 5~Gyr with different timesteps~$dt$, logarithmically spaced between 3~and 60~days. 
The simulations are grouped into bins with 50 simulations each. 
We therefore expect to have on average 0.5 unstable simulations per bin.

\begin{figure}
    \centering
    \resizebox{0.99\columnwidth}{!}{\includegraphics{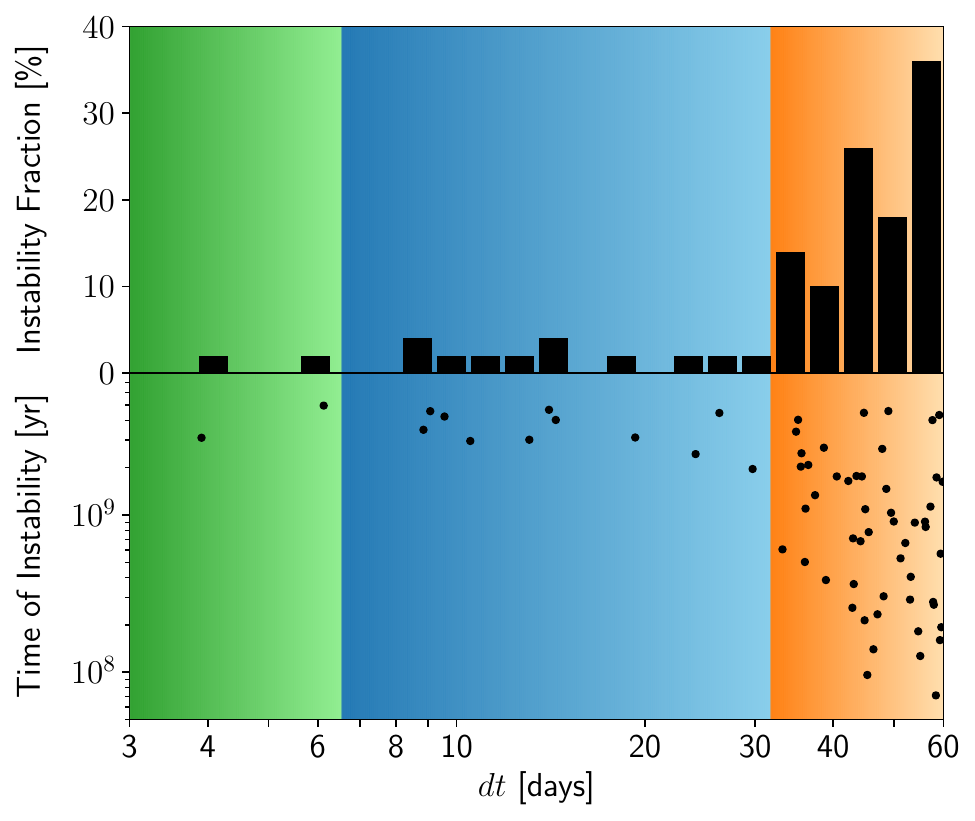}}
    \caption{Top panel: the fraction of simulations that went unstable within 5~Gyr. There are 50 simulations in each timestep bin. 
    Bottom panel: the time of instability (black dots) for individual simulations. 
    The colours indicate the three regimes we identify: statistically converged (green), partially statistically converged (blue), not statistically converged (orange).
    \label{fig:instability-times}
    \vspace{5mm}
    }
\end{figure}
\Needspace{8\baselineskip}
\section{Results}
\Needspace{5\baselineskip}
\subsection{Instability Fraction}\label{sec:insta}
The instability fraction of each bin as well as the instability times of individual simulations are shown in Fig.~\ref{fig:instability-times} as a function of the timestep.
We group the simulations into three regimes. 
Simulations in the green regime with a timestep less than 6.5~days show the expected instability fraction of about 1\%. 
Simulations in the blue regime with a timestep between 6.5 and 32~days have a slightly increased instability fraction of about 2\%. 
Note that in both the green and the blue regime, instabilities happen late, after a few billion years.
Simulations in the orange regime with a timestep larger than 32~days show a qualitatively and quantitatively different picture. 
Instabilities happen much earlier and the instability fraction is much higher.

\begin{figure}
    \centering
    \resizebox{0.99\columnwidth}{!}{\includegraphics{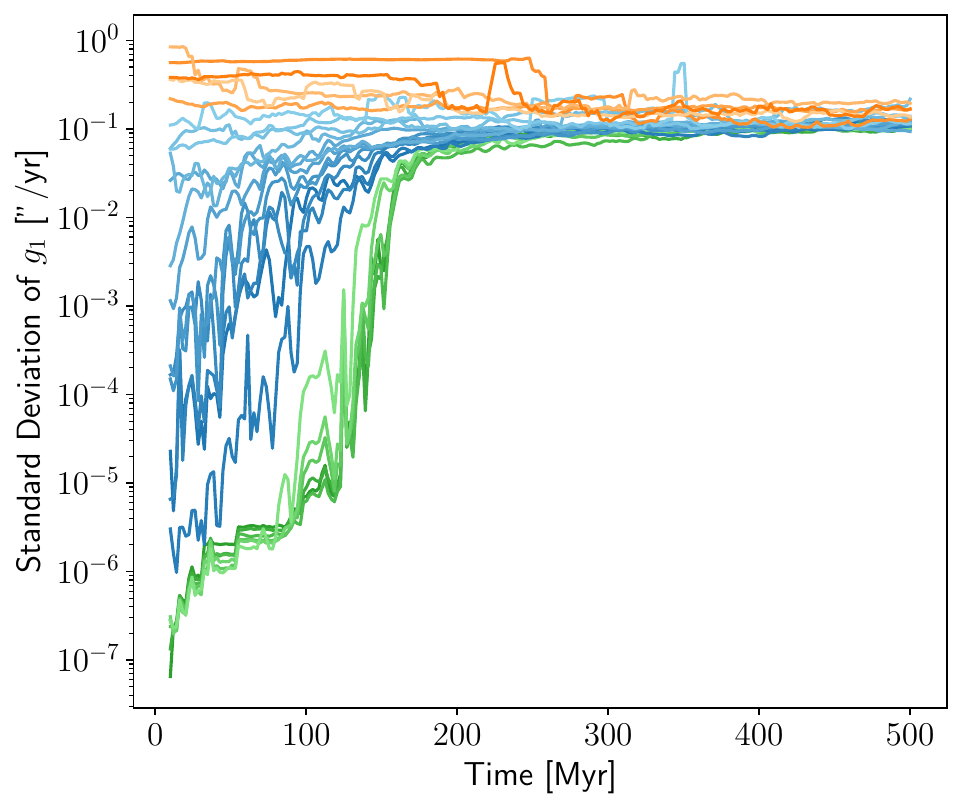}}
    \caption{The standard deviation of the secular frequency $g_1$ as a function of time for different timestep bins. 
    The colours indicate the three regimes and are the same as in Fig.~\ref{fig:instability-times}. 
    \label{fig:std}
    }
\end{figure}
\Needspace{5\baselineskip}
\subsection{Secular Frequency of Mercury} \label{sec:mercury}
For each bin in Fig.~\ref{fig:instability-times}, we also measure the standard deviation of the fundamental frequency $g_1$ as a function of time over the first 500~Myr. 
We compare the standard deviation across runs, rather than the actual values, because the standard deviation provides us with a measure of how fast nearby trajectories diverge.
The $g_1$ secular frequency is of particular interest because it is associated with Mercury, the planet with the shortest orbital period and the highest probability to exhibit large eccentricity variations.
The results are show in Fig.~\ref{fig:std}. The line colours correspond to those in Fig.~\ref{fig:instability-times}.
As one can see in Fig.~\ref{fig:std}, all simulations in the green regime (timesteps less than 6.5~days) show the same exponential divergence.
After the differences become macroscopic, the standard deviations in the different bins level off to the same value,~$\sim0.1\arcsec/$yr.
We therefore consider these simulations to be \textbf{statistically converged}.

Simulations in the blue regime show a different behaviour. 
Within the first 150~Myr, the exponential divergence now clearly depends on the timestep. 
Simulations with a longer timestep diverge faster.
We attribute this to added numerical diffusion \citep{Trani2024}.
These simulations are thus not fully converged according to our definition above. 
However, note that standard deviations of $g_1$ still level off to same value as those in the green bin after $\sim$300~Myr.
We refer to these simulations are \textbf{partially statistically converged}. 

Simulations in the orange regime show once again a very different behaviour. 
There is no exponential divergence visible at all and the standard deviations do not approach the same values as in the simulations of the other regimes. 
Thus, these simulations are \textbf{not statistically converged} and no longer physical. 

We would like to emphasize that the above definition of statistical convergence refers to ensembles of solutions.
The solutions themselves are not converged to the one unique solution whose existence is guaranteed by the Picard–Lindelöf theorem.
Instead, we call the ensembles of solutions statistically converged if one or more observables of the ensembles (e.g. instability rate, rate of divergence of secular frequencies) are statistically indistinguishable from each other.

\Needspace{5\baselineskip}
\subsection{Secular Frequencies after 5~Gyr}\label{sec:after5}
\begin{figure*}
    \centering
    \resizebox{0.99\textwidth}{!}{\includegraphics{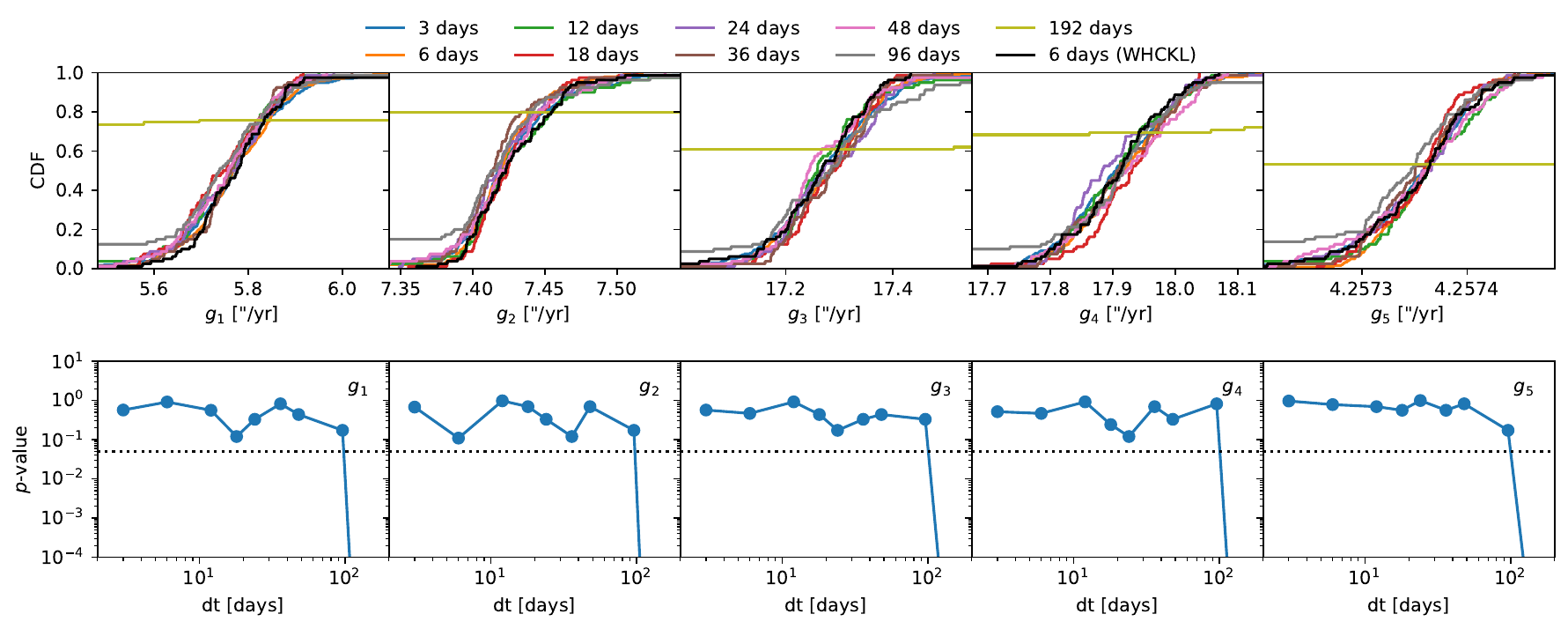}}
    \caption{Top: Cumulative distributions of the secular frequencies $g_1$ through $g_5$ in simulations of the Solar System after 5~Gyr with different timesteps.
    The black line corresponds to integrations using of the high order \texttt{WHCKL} integrator.
    Bottom: p-value as a function of timestep for distributions of $g_1$ through $g_5$. We compare the distributions from each timestep bin to the distribution obtained with WHCKL and a 6~day timestep. The dotted line corresponds to $p=0.05$. 
    \label{fig:sec5gyr}
    }
\end{figure*}
To assess the convergence in the distribution of fundamental frequencies we ran 80 additional simulations for each timestep bin of $3, 6, 12, 18, 24, 36, 48, 96,$~and $192$~days.
For these runs, we use the standard \texttt{WHfast} integrator of order $O(\epsilon dt^{2})$ without symplectic correctors. 
We also run one set of 80 simulations with a 6~day timestep using the \texttt{WHCKL} integrator.
In contrast to the previous runs, we do not stop these simulations even if a close encounter or an ejection occurs.
After 5~Gyr, we integrate all simulations for an additional 10~Myr with a 6~day timestep to accurately measure the fundamental frequencies with a modified Fourier transform \citep{Laskar1988,Laskar1990} for which we use the implementation of \cite{Sidlichovsky1996}.

The secular frequencies $g_1$ (Mercury) through $g_5$ (Jupiter) at the end of the 5~Gyr integrations are shown as cumulative distributions in in the top row of Fig.~\ref{fig:sec5gyr}.
Simulations in which an instability occurred are included in the distributions. 
For those the fundamental frequencies are vastly different and thus might appear outside of the bounds of the plots.
The frequencies $g_6$, $g_7$, and  $g_8$ associated with Saturn, Neptune, and Uranus are not plotted but show a very similar behaviour.

We also perform a Kolmogorov–Smirnov test \citep{Kolmogorov1933,Smirnov1948} by comparing the distribution for each timestep bin with those of the WHCKL runs which we consider having the highest accuracy (see next section).
The $p$-values and the commonly used threshold of $p=0.05$ are plotted in the bottom row of Fig.~\ref{fig:sec5gyr}.
One can see that the distributions show no statistically significant difference ($p$-value smaller than 0.05) in this late-stage metric for simulations that use a timestep of 96~days or less.
This further justifies our nomenclature of labelling these simulations statistically or partially statistically converged.
However, note that we saw a difference in the instability rate for timesteps larger than 32~days (Sect.~\ref{sec:insta}).
This late-stage agreement in secular frequencies is thus only a necessary but not sufficient criteria to ensure that results are physical.

\Needspace{5\baselineskip}
\subsection{Energy Errors after 5~Gyr} \label{sec:energy}
\begin{figure}
    \centering
    \resizebox{0.99\columnwidth}{!}{\includegraphics{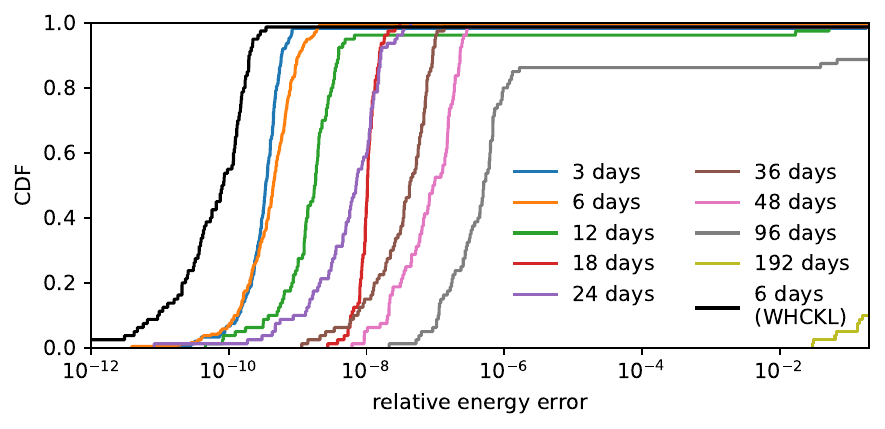}}
    \caption{
        Cumulative distribution of relative energy errors at the end of the 5~Gyr integration for each timestep bin. 
    \label{fig:energy}
    }
\end{figure}
\begin{figure}
    \centering
    \resizebox{0.99\columnwidth}{!}{\includegraphics{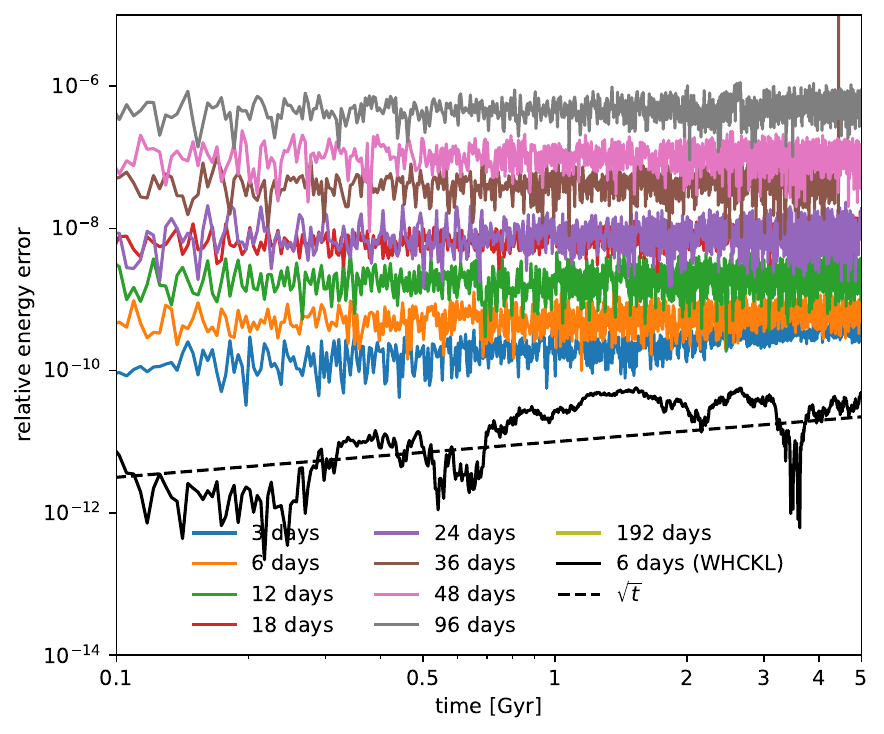}}
    \caption{
        The relative energy error as a function time for a randomly chosen simulation from each timestep bin.
        Also plotted is a line proportional to the square root of time which is the expected time dependence for a simulation that follow Brouwer's law.
    \label{fig:energy1}
    }
\end{figure}
We also measure the relative energy error $|\Delta E/E|$ at the end of the 5~Gyr simulations described in Sect.~\ref{sec:after5}. 
The cumulative distributions for each timestep bin are plotted in Fig.~\ref{fig:energy}.
One can see that the simulations using WHCKL and a 6~day timestep have the lowest median relative energy error of less than $10^{-10}$. Simulations with a 3~day timestep have a median energy error of $\sim3\cdot10^{-10}$.
For larger timesteps the energy error then increases proportionally to $dt^2$. 
For simulations with a 96~day timestep it reaches $\sim5\cdot10^{-7}$.
The median error  for simulations with a timestep of 192~days is of order unity as it is dominated by the large number of unstable systems.

The time evolution of one randomly chosen simulation from each timestep bin is plotted in Fig.~\ref{fig:energy1}.
One can see that the energy error remains approximately constant during the integration as expected for a symplectic integrator.
The energy error of the WHCKL simulation with a 6~day timestep shows a growth $\sim\sqrt{t}$.
This is the expected behaviour for a simulation that follows Brouwer's law \citep{Brouwer1937,ReinTamayo2015} where random errors due to finite floating point precision accumulate in an unbiased way.
This is the best error behaviour we can achieve with floating point arithmetic.

Monitoring the energy error is a necessary requirement to ensure physical results.
However, note that it is by no means sufficient: aside from the 192~day bin, the energy errors are bound and in general look well behaved for all timestep bins.
With only this data at hand, one could come to the false conclusion that simulations with a 96~day timestep and a relative energy error of less than $10^{-6}$ are capturing the dynamical evolution of the system well.
But this is clearly not the case as can be seen by looking at the instability rate presented in Sect~\ref{sec:insta}.

\Needspace{5\baselineskip}
\section{Discussion}
Our results show that even simulations with very large timesteps -- up to about a~third of Mercury's period (!) -- can capture the chaotic diffusion phase of the Solar System surprisingly well~(Sects.~\ref{sec:mercury} and~\ref{sec:after5}). 
The dynamical picture described in Sect.~\ref{sec:secdyn} allows us to explain why the instability in the partially statistically converged (blue) regime is only slightly increased compared to the statistically converged (green) regime (Sect.~\ref{sec:insta}). 
The partially statistically converged simulations clearly do not reproduce the initial exponential divergence.
This is due to the added numerical diffusion which is dependent on the timestep \citep{Trani2024}.
Because of this non-physical diffusion, the Lyapunov time of the system differs.
Yet, our numerical results show that these simulations still more or less accurately capture the chaotic diffusion which dominates over long timescales. 
The slightly higher instability rate in partially statistically converged simulations is consistent with this interpretation as the added numerical diffusion can help push the system into otherwise harder to reach areas of parameter space which then lead to an instability.
The fact that all the instabilities in the partially statistically converged regime happen late further supports our argument that these instabilities are driven by chaotic diffusion, now part physical, part numerical, but in a very similar way to those in the fully statistically converged regime.

Simulations which are not statistically converged (orange) on the other hand fail to correctly capture both the initial exponential divergence and the later chaotic diffusion. 
The high rate of instabilities, the early instability times, and the different distribution of fundamental frequencies at late times are clear indications that a different mechanism is driving these system towards instability.
In addition to the increased numerical diffusion, \cite{Rauch1999} showed that large timesteps can also lead to so-called timestep resonances.
The timescales associated with timestep resonances are significantly shorter than those associated with secular resonances, thus explaining the increased instability rate and the much early onset of instabilities. 

\texttt{WHFast} is a second order scheme and we find that the energy error indeed scales as $|\Delta E/E| \propto dt^2$ (Sect.~\ref{sec:energy}).
Whereas this is a confirmation that the scheme is implemented correctly and is a necessary requirement to obtain physical results, it provides little information as to whether the physical results from an ensemble of simulations are statistically converged or not. 
Specifically the relative energy error does not show the qualitative differences that we observe between the statistically converged, partially statistically converged, and not statistically converged regimes.

\Needspace{5\baselineskip}
\section{Conclusions}
In this paper we presented results from several thousand 5~Gyr integrations of the Solar System with different timesteps.
Our results show that simulations with timesteps of 6.5~days or less are statistically converged.
For these simulations, no statistically significant differences can be observed even after 5~Gyr.

Surprisingly, we find that even simulations with timesteps as large as 32~days are statistically almost indistinguishable after 5~Gyr in terms of instability fraction, time of instability, and secular frequencies.
Such partially statistically converged simulations might give unreliable results over short timescales, but the chaotic diffusion in the Solar System comes to our rescue over longer timescales. 
The diffusion washes out any information about the initial conditions. 
Any error we introduce because of the large timestep doesn't matter as long as it remains small compared to the physical effects driving the planets' evolution. 

This argument shouldn't come as a surprise because, regardless of the timestep we use, we always introduce some amount of numerical noise ($10^{-16}$) into the simulation at every mathematical operation due to finite floating point precision. 
The novel result from this work is that we were able to show that numerical artifacts several order of magnitude larger than floating point precision and originating from a finite timestep do also not affect the planets' long-term evolution.

Note that our goal was to determine the probability of orbits changing significantly.
The specific criteria we chose (see Sect.~\ref{sec:setup}) looks at whether two planets enter their mutual Hill sphere\footnote{See also Table~1 of the Supplementary Information of \cite{LaskarGastineau2009} which compares different instability criteria.}.
If our goal were to resolve actual physical collisions, our results would change significantly as none of the integrations with a fixed timestep of a few days is able to resolve such events.

Previous studies \citep{Wisdom2015,Hernandez2022} have argued that one might need up to 20 timesteps per pericenter passage time to avoid timestep resonances and achieve reliable results. 
This corresponds to a timestep of less than 3 days for the Solar System. 
\cite{Wisdom2015} identifies potentially problematic timestep resonances, but looked only at relatively short term integrations (thousands of orbits) and very clean test cases (the Stark problem\footnote{The Stark problem is a two body problem where one of the bodies has an additional acceleration of constant magnitude and direction, see e.g. \cite{Lantoine2011}.} with one planet and a system with two giant planets). 
Whereas these test cases contain mean motion resonances, they do not contain the kind of secular resonances that are driving the long-term evolution of the Solar System \citep{Mogavero2022}.
The crucial difference is the timescale.
For planets with masses similar to those of the (inner) Solar System, secular timescales are at least two orders of magnitude longer than those of mean motion resonances \citep{TamayoHadden2025}.
Thus any potential timestep resonance in the Solar System will be much weaker than in a system with strong mean motion resonances because the secular timescale and the timestep are so well separated.
Another difference between such clean test cases and an integration of the full Solar System is the number of planets.
For example, in contrast to the Stark problem, the many different planet-planet interactions in the Solar System constantly change the orbits of planets and secular frequencies.
Because of that the Solar System frequently enters a secular resonance only to shortly afterwards leave it again \citep{LithwickWu2011}.
So even if at one instance, a timestep resonance occurs, the system might not be in the resonance long enough to be significantly affected by it.
As this paper only looks at data from Solar System integrations, a primarily secularly evolving system, the results should not be blindly applied to other systems, especially those with strong mean motion resonances. 

An important point to raise here is that all integrations in this paper were performed with a symplectic integrator.
Because of chaotic diffusion, the Solar System is undergoing a random walk. 
The variations due to this random walk reach order unity on a short timescale of only a few 100~Myr (see Fig.~\ref{fig:sec5gyr}).
Thus there is no need to get every individual integration exactly right.
Several times during a 5~Gyr integration, we loose any information about both the initial conditions and any potential numerical error we made.
One only needs to make sure that numerical artifacts (e.g.\ timestep resonances, finite floating point precision, numerical diffusion) do not dominate over physical effects (e.g.\ mean-motion resonances, secular resonances, chaotic diffusion).
This is further supported by the fact that the \texttt{WHCKL} integrations provide identical results compared to the plain \texttt{WHFast} integrations, despite the fact that \texttt{WHCKL} is more accurate in short term integrations \citep{ReinTamayoBrown2019}. 
We note however that it is important that any superimposed numerical effect doesn't lead to long-term drifts \citep{Trani2024}.
This is the reason why symplectic integrators and unbiased implementations are crucial for this task \citep{Brouwer1937,ReinTamayo2015}. 
The conclusion from this paper is that numerical artifacts are negligible for Solar System integrations if the timestep is 6.5~days or less.
Even timesteps as large as 32~days only increase the instability rate by a factor of 2 over 5~Gyr because chaotic diffusion still mostly dominates over numerical artifacts.

We find that the historically often cited metric of the relative energy error can only serve as a necessary condition to ensure that nothing has gone catastrophically wrong, but cannot be used to reliably determine whether physical results of simulations are statistically converged or not.
Instead, we strongly recommend that researchers always verify that the physical results they actually care about are independent of the timestep used.
This is of course also not a sufficient condition to ensure statistical convergence: all tested timesteps might be too large.
However, it is another necessary condition which can provide valuable insights.
And as we have shown in this paper, a surprisingly large timestep can result in converged results. 
Thus, a computationally inexpensive way to gather further evidence for statistically converged results is to rerun a simulation with an increased timestep (which takes less time than the original simulation).
If the results change, then additional simulations with a smaller timestep should be run.

\Needspace{5\baselineskip}
\section*{Software}
The numerical integration methods underlying this article are part of the REBOUND package, available at \url{https://github.com/hannorein/rebound}.
The code for the modified Fourier transform is available at \url{https://boulder.swri.edu/~davidn/fmft/fmft.html}.
This research was made possible by the open-source projects 
\texttt{Jupyter} \citep{jupyter}, \texttt{iPython} \citep{ipython}, and  \texttt{matplotlib} \citep{matplotlib, matplotlib2}.

\Needspace{5\baselineskip}
\section*{Acknowledgments}
We thank Daniel Tamayo, Nathan Kaib, and Sean Raymond for helpful discussions and feedback on this paper.
We thank two anonymous referees whose constructive feedback significantly improved the manuscript.
This research has been supported by the Natural Sciences and Engineering Research Council (NSERC) Discovery Grants RGPIN-2014-04553 and RGPIN-2020-04513.
This research was enabled in part by support provided by Digital Research Alliance of Canada (formerly Compute Canada; \href{https://alliancecan.ca/en}{alliancecan.ca}).
Computations were performed on the Niagara supercomputer \citep{SciNet2010, Ponce2019} at the SciNet HPC Consortium (\href{www.scinethpc.ca}{scinethpc.ca}). 
SciNet is funded by the following: the Canada Foundation for Innovation; the Government of Ontario; Ontario Research Fund -- Research Excellence; and the University of Toronto.

\Needspace{5\baselineskip}
\bibliography{full}
\bibliographystyle{mnras.bst}

\label{lastpage}
\end{document}